\documentclass[iop, apjl]{emulateapj}
\usepackage{amsmath}
\usepackage{mathrsfs}

\shortauthors{Rappazzo  \& Parker}

\begin{document}

\title{Current Sheets Formation in Tangled Coronal Magnetic Fields}

\author{A. F. Rappazzo$^{1,2}$ and E.~N.~Parker$^3$}
\affil{
$^1$Bartol Research Institute, Department of Physics and Astronomy,
University of Delaware, Newark, DE 19716, USA\\ 
$^2$Advanced Heliophysics, 1127 E Del Mar Blvd, Suite 425, Pasadena,
CA 91106, USA\\ 
$^3$Enrico Fermi Institute, University of Chicago, Chicago, IL 60637, USA}

\begin{abstract}
We investigate the dynamical evolution of 
magnetic fields in closed regions of solar and stellar coronae.
To understand under which conditions current sheets form,
we examine dissipative and ideal reduced 
magnetohydrodynamic models in cartesian geometry,
where two magnetic field components are present:
the strong guide field $B_0$, extended along the axial direction,
and the dynamical orthogonal field $\mathbf{b}$.
Magnetic field lines thread the system along the axial 
direction, that spans the length $L$, and are line-tied 
at the top and bottom plates.
The magnetic field $b$ initially has only large scales, 
with its gradient (current) length-scale of order $\ell_b$.
We identify the magnetic intensity threshold $b/B_0 \sim \ell_b/L$.
For values of $b$ below this threshold, field-line tension inhibits the formation
of current sheets, while above the threshold they form quickly
on fast ideal timescales. In the ideal case, above the magnetic threshold,
we show that  current sheets thickness decreases in time until 
it becomes smaller than the grid resolution,
with the analyticity strip width $\delta$ decreasing at least exponentially,
after which the simulations become under-resolved.
\end{abstract}


\section{Introduction}

All late type main sequence stars, for which the 
Sun is the prototype, emit X-rays \citep{gud04}. And
solar observations at increasingly higher resolutions show
that the X-ray corona has structures at all resolved
scales \citep{cgw13}.

Convective motions, that
have more than enough energy to heat the corona 
at temperatures $>10^6\, \textrm{K}$,
shuffle continuously the coronal magnetic field line footpoints,
giving rise to a magnetic field that is not in equilibrium \citep{park72, park00, vb85}.
\citet{park72, park88, park94, park12} 
pointed out that
\emph{current sheets} are an intrinsic part of the final equilibrium of
almost all interlaced field line topologies.  So the asymptotic relaxation
of the interlaced field to equilibrium necessarily involves the formation
of current sheets, providing energy dissipation presumably concentrated
in small impulsive heating events, so-called \emph{nanoflares}.
This picture has had a strong impact on 
the thermodynamical modeling of the closed corona
\citep{klim06}, but  it is still controversial if and under 
which circumstances current sheets form.

Analytical models \citep{vb85, ant87, cls97} claim that in general well-behaved
photospheric motions will not lead to the formation of current sheets,
and that only a discontinuous velocity field can form discontinuities
in the coronal magnetic field, and counterexamples of
well-behaved solutions of the magnetostatic equations
have been reported \citep{rk82, bogo00}.

Alternatively \cite{vb86}  proposed that the random character of 
footpoint motions might generate,
on time-scales much longer than photospheric convection time-scales,
uniformly distributed small-scale current layers that would heat the
corona without forming discontinuous structures.

Numerical simulations of boundary forced models
\citep{evpp96, dg97, rved07}  suggest
that the nonlinear dynamics of this system can 
be modeled as a magnetically dominated instance
of magnetohydrodynamic (MHD) turbulence, implicitly
implying that current sheets thickness is limited only by
numerical diffusion (i.e., resolution) in dissipative MHD.
But recent simulations of the decay of an initially 
braided magnetic configuration \citep{wshp09} have shown 
that in some instances the system forms only 
large-scale current layers of thickness 
much larger than the resolution scale, 
in stark contrast with the recent result
supporting the development of finite time singularities
in the cold plasma regime \citep{low13}.

Furthermore, recent investigations 
suggest that the rate of magnetic reconnection
can be very fast in low collisional plasmas,
both in the MHD  \citep{lv99, lap08, lus09, hb10}
and the collisionless regime \citep{sdr99}.
Therefore, in order to have an X-ray corona, it is \emph{critical}
that current sheets form, but only \emph{above} a
magnetic energy  threshold.
Indeed the energy flux injected in the corona by 
photospheric motions is
the average Poynting flux 
$\langle S_z \rangle = B_0\, \langle \mathbf{u}_\textrm{ph} \cdot \mathbf{b} \rangle/4\pi$ 
\citep[][\S 3.1]{rved08}, 
that depends not only on the photospheric velocity $\mathbf{u}_\textrm{ph}$
and the axial guide field $B_0$,
but also on the dynamic magnetic field $\mathbf{b}$,
and if dissipation keeps low the value of $b$, the flux $\langle S_z \rangle$ will be
too low to sustain the corona \citep{wn77}. 

In this letter we investigate, in a cartesian model of the closed 
corona, under which conditions current sheets form,  and their 
dynamical properties.

\section{Model}

A closed region of the solar corona is modeled in 
cartesian geometry as a plasma with uniform 
density $\rho_0$ embedded in a \emph{strong  and
homogeneous axial magnetic field} 
$\mathbf{B}_0 = B_0\, \mathbf{\hat{e}}_z$
well suited to be studied \citep[e.g., see][]{derv12}, 
as in previous work, with the equations
of  reduced magnetohydrodynamics (RMHD).
Introducing the velocity and magnetic field potentials $\varphi$ and $\psi$, for which
$\mathbf{u} = \nabla \times \left( \varphi\, \mathbf{\hat{e}}_z  \right)$,
$\mathbf{b} = \nabla \times \left( \psi\, \mathbf{\hat{e}}_z  \right)$, vorticity
$\omega = - \nabla^2_{\!_{\!\perp}} \varphi$, and the current density 
$j = - \nabla^2_{\!_{\!\perp}} \psi$, the nondimensional RMHD equations \citep{kp74, str76}
are:
\begin{eqnarray}
&& \partial_t \psi = \left[ \varphi, \psi \right] 
+ B_0\, \partial_z \varphi
+ \eta_{_n} \nabla_{_{\!\!\perp}}^{2n} \psi, \quad \label{eq:eq1} \\
&& \partial_t \omega = \left[ j, \psi \right] - \left[ \omega, \varphi \right]
+ B_0\, \partial_z j
+ \nu_{_{\!n}} \nabla_{_{\!\!\perp}}^{2n} \omega, \label{eq:eq2}
\end{eqnarray}
The Poisson bracket of functions $g$ and $h$ is defined as
$[g,h] = \partial_x g\, \partial_y h - \partial_y g\, \partial_x h$
(e.g., $[j,\psi]=\mathbf{b} \cdot \nabla j$), and
Laplacian operators have only orthogonal components.
To render the equations nondimensional we have first expressed the 
magnetic fields as an Alfv\'en velocity ($b \rightarrow b/\sqrt{4\pi \rho_0}$)
and then normalized all velocities to $u^{\ast} = 1$~km\,s$^{-1}$, typical
value for the photosphere.
The domain spans $ 0 \le x, y, \le \ell$, $ 0 \le z \le L$, 
with $\ell = 1$ and $L=10$.
Magnetic field lines are line-tied to a motionless photosphere 
at the top and bottom plates $z=0,10$, where a velocity $\mathbf{u} = 0$
is imposed. In the perpendicular ($x$-$y$) directions
we use a  pseudo-spectral scheme with
periodic boundary conditions, and along $z$
a second-order finite difference scheme.
For a more detailed description of 
the model and numerical code see \cite{rved07, rved08}.

Dissipative simulations use hyper-diffusion \citep{bis03}, that effectively
limits diffusion to the small scales, with $n=4$ and 
$\nu_n = \eta_n = \left(-1\right)^{n+1} / R_n$ 
($R_n$ corresponds to the Reynolds number for $n=1$) 
\cite[see][\S 2.1]{rved08}, while ideal simulations
implement $\nu_n = \eta_n = 0$.

\subsection{Initial conditions}

Simulations are started at time $t=0$ with a vanishing
velocity  $\mathbf{u} = 0$ everywhere, and a uniform
and homogeneous guide field $B_0$.
The orthogonal field $\mathbf{b}$ consists of 
\emph{large-scale} Fourier modes, set expanding
the magnetic potential  in the following way:{\small
{\setlength\arraycolsep{-15pt}
\begin{eqnarray}
&& \psi_0  = b_0
\sum_{rsm} ( 2 \mathscr{E}_{_m})^{\frac{1}{2}}\ \frac{\alpha_{rsm} \ \sin \left( \mathbf{k}_{rsm}\! \cdot \mathbf{x} + 2\pi \xi_{rsm} \right) }
{ k_{rs}\ \sqrt{ \sum_{ij} \alpha^2_{ijm} } }  
\label{eq:pot4}\\
&&\textrm{with} \quad \mathbf{k}_{rsm} = \frac{2\pi}{\ell}
\left( r\, \mathbf{\hat{e}}_x + s\, \mathbf{\hat{e}}_y \right)
+\frac{2\pi}{L} m\, \mathbf{\hat{e}}_z, \nonumber \\
&& \textrm{and} \quad k_{rs} = \frac{2\pi}{\ell} \sqrt{ r^2 + s^2 }, \nonumber
\end{eqnarray}
}}where the coefficients $\alpha_{rsm}$ and $\xi_{rsm}$ are two independent 
sets of random numbers uniformly distributed between 0 and 1.
The orthogonal wave-numbers $(r,s)$  are always
in the range $3 \le (r^2+s^2)^{1/2} \le 4$, 
while the parallel amplitudes $\mathscr{E}_m$  (with $\sum_m \mathscr{E}_m =1$)
are set to distribute the energy in different ways in the axial direction.
Given the orthogonality of the base used in Eq.~(\ref{eq:pot4})
the normalization factors guarantee that
for any choice of the amplitudes the rms of the
magnetic field is set to 
$ b = \langle  b_x^2 + b_y^2  \rangle^{1/2} = b_0$,
while for total magnetic energy 
$E_{_{\!M}} = b_0^2 V/2 \sum_m \mathscr{E}_m$,
i.e., $\mathscr{E}_m$ is the fraction of magnetic energy in the 
\emph{parallel} mode~$m$.
Two-dimensional (2D) configurations invariants along $z$  are obtained 
considering the single mode $m=0$ with $\mathscr{E}_0=1$.

\subsection{Equilibria} \label{sec:eq}

Neglecting velocity and diffusion, equilibria of 
Eqs.~(\ref{eq:eq1})-(\ref{eq:eq2}) are given by
\begin{equation} \label{eq:eqb}
\mathbf{b} \cdot \nabla j + B_0 \partial_z j = 0 \quad \rightarrow \quad
\partial_z j = - \mathbf{b}/B_0 \cdot \nabla j.
\end{equation}
This equation has the same structure of the 
2D Euler equation for vorticity \citep{vb85},
as can be verified substituting $(\mathbf{b}/B_0,\ j/B_0) \rightarrow (\mathbf{u},\ \omega)$,
\emph{and $z$ with time $t$}, and
has been studied extensively in 2D hydrodynamic turbulence \citep{km80}.
Given a smooth $j$ at any $x$-$y$ plane it admits a
unique and regular solution \citep[not singular,][]{rs78}, with an
asymmetric structure along $z$, as
$\bf b$ acquires larger scales
through an inverse cascade, and currents
are stretched (with $j$ constant along the ``fluid element'').
\cite{park00,park12} points out that the observed \emph{disordered} 
photospheric motions will in general induce an interlaced coronal 
magnetic field that does not have this structure and
will thus generally \emph{not be in equilibrium}.

Given the long time-scale of photospheric convection
respect to the fast Alfv\'en crossing time the 
coronal structure will be dominated by the low-frequency $m=0$ mode.
In this 2D limit the equilibrium condition (\ref{eq:eqb}) reduces to
$\mathbf{b} \cdot \nabla j =0$, i.e., 
$j$ is constant over the field lines of  $\mathbf{b}$,
a configuration too symmetric to occur in  coronal fields.

We therefore use as initial condition a magnetic field that
is \emph{not in equilibrium}, obtained
using random amplitudes in our initial conditions (\ref{eq:pot4})
(a 2D example is shown in Figure~\ref{fig:fig5}, $t=0$).

\section{Results}

We first consider initial conditions with $m=0$ 
(\emph{not in equilibrium}, with field lines of $\mathbf{b}$
same as in Figure~\ref{fig:fig5} at $t=0$), 
invariant along the
axial direction ($\partial_z = 0$), and $b_0 = 0.1 B_0$, with $B_0=10^3$.
To understand the effect of line-tying we perform 
two \emph{dissipative} sets of simulations with same parameters but different 
boundary conditions along $z$:
\emph{periodic} (with $R_4 = 3\cdot 10^{20}$ and a $1024^2 \times 512$ grid),  
and \emph{line-tied} (with $R_4 = 10^{19}$ and $512^2 \times 256$ grids).

The \emph{periodic case} is the limit of a very long loop, when line-tying
boundary conditions do not have a strong bearing on the dynamics.
In this case the system is approximately invariant along $z$, 
and the solution will be two-dimensional as the initial condition ($\partial_z = 0$).
This configuration is akin to 2D turbulence decay \citep{bis03, hos95},
except that the initial velocity vanishes.

\begin{figure}
\includegraphics[scale=.33]{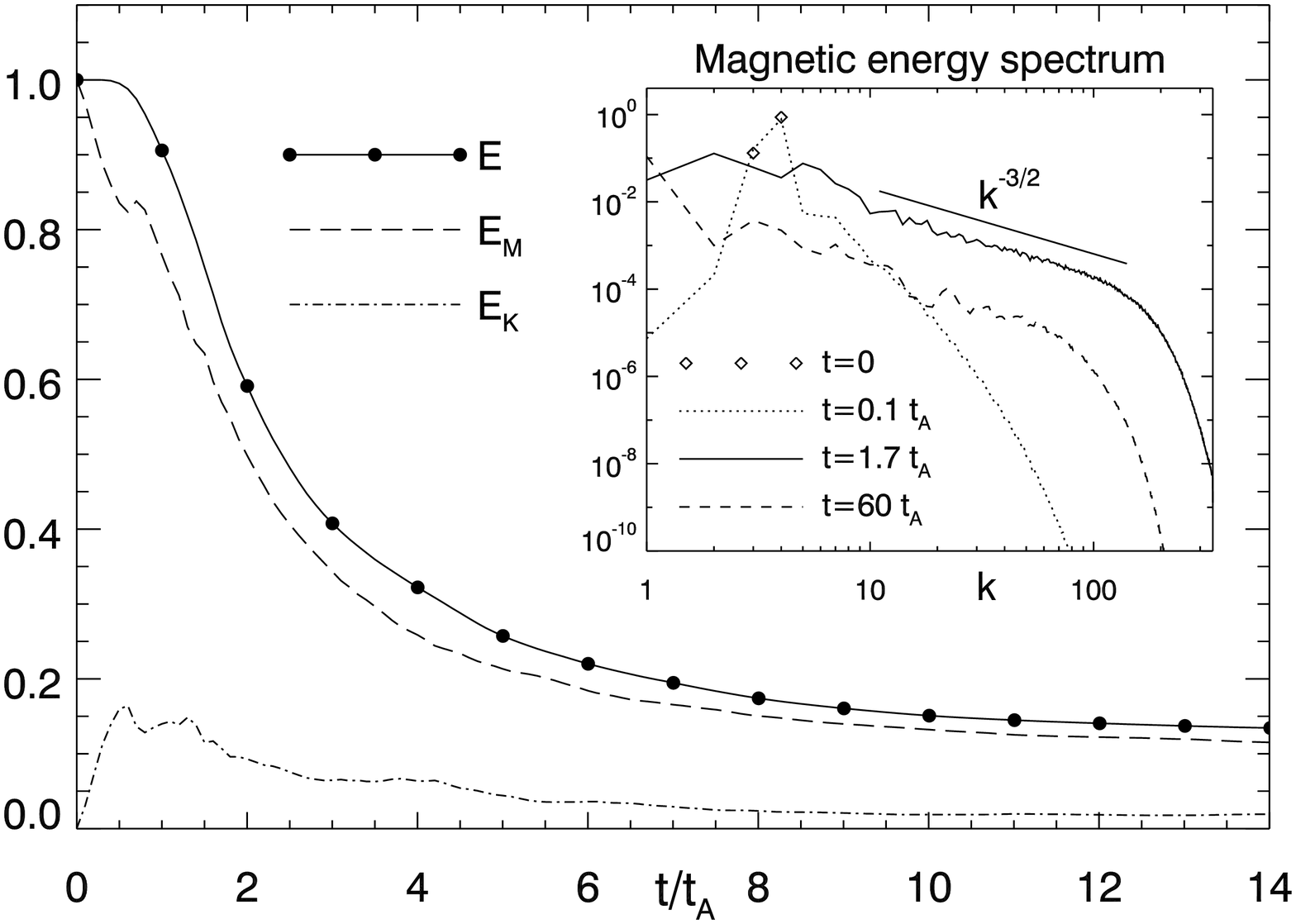}
\caption{Periodic simulation with 2D initial conditions:
Magnetic $E_M$, Kinetic $E_K$ and total energies 
$E$ versus time (normalized with the initial total energy). 
Magnetic energy spectra at selected
times are shown in the inset.
\label{fig:fig1}}
\end{figure}

At $t=0$ the magnetic field is not in equilibrium, thus no instability develops,
but the non-vanishing Lorentz force transfers $\sim 15\%$ 
of magnetic energy $E_M$ into kinetic energy $E_K$
(with $E_M$ bigger than $E_K$ in all simulations), while
until time $t\sim 0.3\, \tau_{A}$ total energy $E$ is conserved
(Figure~\ref{fig:fig1}).
Subsequently magnetic energy, initially present only in \emph{perpendicular} modes $k=3$ and $4$,
cascades in Fourier space toward higher wavenumbers 
(Figure~\ref{fig:fig1}, inset, $E =\sum_k E_k$),
corresponding to the formation of current sheets in physical space.
At the peak of dissipation ($t \sim 1.7\, \tau_A$) the spectrum
exhibits a $k^{-3/2}$ power-law and is fully extended toward 
the maximum wavenumber ($k_\textrm{max} = 341$).

\begin{figure}
\includegraphics[scale=.33]{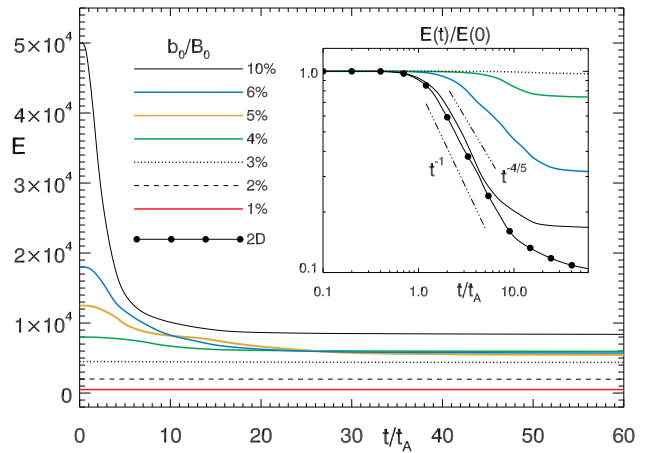}
\caption{Total magnetic energy versus time (in logarithmic
scale in the inset) for line-tied simulations with different
values of $b_0/B_0$, and the 2D simulation with 
$b_0/B_0 = 10\%$.
\label{fig:fig2}}
\end{figure}

Once current sheets are formed dissipation occurs,
total and magnetic energies decay approximately 
with the power-law $E \propto t^{-4/5}$ 
(see inset in Figure~\ref{fig:fig2}) as in the 2D turbulence case \citep{gpp97}, 
while kinetic energy decays as $E_K \propto t^{-1}$ before vanishing asymptotically.  
The magnetic field loses energy at high
wavenumbers (Figure~\ref{fig:fig1}, inset $t=60\, \tau_A$), thus 
current sheets disappear,
and the system relaxes to a stage with $\sim 5\%$ of the initial magnetic
energy and a very small velocity.
At the same time an \emph{inverse cascade} occurs, 
transferring energy at the largest scales (particularly in mode $k=1$),
so that the asymptotic state consists of large-scale magnetic islands, with
large-scale current layers and no current sheets,
and this process can be described as the 2D analog 
of Taylor relaxation \citep{tay86}.

For this 2D case (periodic boundary conditions, with $\partial_z = 0$) 
given a solution of Eqs.~(\ref{eq:eq1})-(\ref{eq:eq2}) 
with initial condition $b=b_0$, solutions with $b= \sigma b_0$ and same
random amplitudes are self-similar in time\footnote{
Strictly speaking these self-similar solutions would require 
the Reynolds number to scale as $R_{\alpha} = \alpha R$, 
but in the high-Reynolds regime the solutions of decaying 
turbulence do not depend on the Reynolds number \citep{bis03, gpp97}. 
}:
$\psi_{\sigma} (\mathbf{x},t) = \sigma\, \psi_{0} (\mathbf{x}, \sigma\, t)$.
Consequently all these solutions have a similar structure and the
temporal evolution differs only for the scaling factor $\sigma$.
In particular if current sheets form for a certain value of $b_0$,
they will always do for any value of $b_0$ at scaled times.
Analogously energy will exhibit a power-law decay with the \emph{same
exponent} as $E_\sigma (t) = \sigma^2 E_0 (\sigma t)$.

When the same initial condition is used with \emph{line-tying} boundary 
conditions, the system is no longer invariant along $z$, as now the
velocity must vanish at the top and bottom plates $z=0,L$, therefore
the velocity cannot develop uniformly along z as in the periodic case.

The temporal evolution of total energy for line-tied
simulations with different values of $b_0$ is shown in 
Figure~\ref{fig:fig2}. While the dynamics of the system
with $b_0/B_0 = 10\%$ is similar to the 2D 
case with energy dissipating $\sim 80\%$ of its initial value, 
the behavior is increasingly different for lower
values of $b_0$, with less energy getting dissipated. 
For $b_0/B_0 \lesssim 3\%$ no significant
energy dissipation nor decay are observed,
and also for the decaying cases their dynamics are quenched 
once energy crosses this threshold.
As shown in the inset in Fig.~\ref{fig:fig2} 
energies decay with different power-law indices,
indicating that self-similarity is lost and new
dynamics emerge.

There are therefore two \emph{antagonistic} forces at work.
The system starts to behave as in the 2D case,
with the tension of perpendicular field lines 
$\mathbf{b} \cdot \nabla \mathbf{b}$ creating an orthogonal velocity,
that coupled with all others \emph{nonlinear terms} are the only ones that 
can cascade energy and generate current sheets.
But this displaces the total line-tied (axially directed) field lines,
and  is then opposed by the axial tension $B_0 \partial_z \mathbf{b}$
that resists bending, and together with the other \emph{linear term} $\propto B_0 \partial_z$  
tends to impose the vanishing boundary velocity in the whole box. 
Furthermore the pattern of the  boundary velocity does not match that of the
velocity generated by the nonlinear terms in Eqs.~(\ref{eq:eq1})-(\ref{eq:eq2})
(also for $\mathbf{u}_\textrm{ph} \neq 0$). Consequently line-tying opposes current sheet
formation, more efficiently the smaller the value of $b/B_0$.

\begin{figure}
\includegraphics[scale=.4]{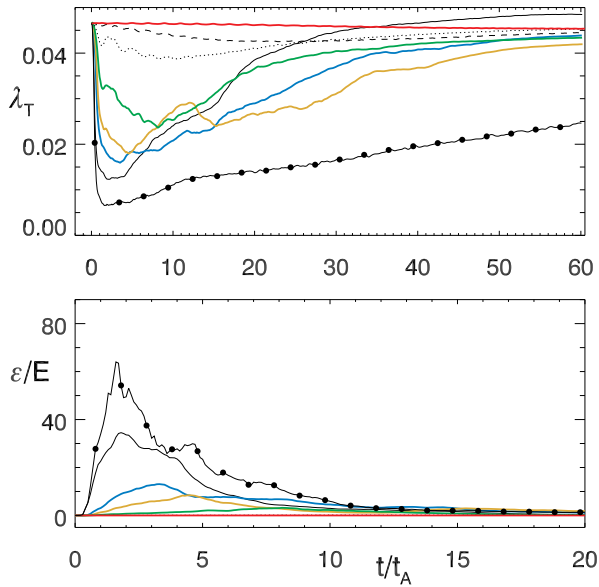}
\caption{Taylor microscale $\lambda_T$ and normalized energy dissipation
rate $\epsilon/E$ ($\epsilon=dE/dt$) versus time.
Color codes same as in Figure~\ref{fig:fig2}.
\label{fig:fig3}}
\end{figure}

For a low ratio of $b_0/B_0$ the axial field line tension
dominates and impedes the formation of current sheets.
This is quantitatively shown in Fig.~\ref{fig:fig3}.
Magnetic Taylor microscale 
$\lambda_T = ( \langle b^2 \rangle / \langle j^2 \rangle  )^{1/2}$
measures the average length-scale of magnetic gradients \citep{mdw05}.
The smallest scales are reached in the 2D simulation, while
for the line-tied case the minimum value of $\lambda_T$
increases with $b_0/B_0$, but  for $b_0/B_0 < 4\%$
no significant gradients are formed. 
While the 2D case retains larger gradients in the asymptotic
state, line-tying sharply removes small-scales after the dissipative
peak.
At the same time normalized energy dissipation rate 
$\epsilon/E$ ($\epsilon = dE/dt$)
decreases sharply for lower values of $b_0/B_0$, with
the 2D case reaching a higher dissipative peak. 

We have performed similar sets of simulations with different initial
conditions,  including more modes besides $m=0$, 
and they show a similar behavior to
that shown in Figs.~\ref{fig:fig1}-\ref{fig:fig3} and will be described
in detail in an upcoming paper.

We conclude that current sheets form when the orthogonal
Lorentz force $\mathbf{b} \cdot \nabla \mathbf{b}$ is stronger
than the field line tension term $B_0 \partial_z \mathbf{b}$.
From initial condition (\ref{eq:pot4}) we can estimate 
the gradient lenght-scale of $b$ in the orthogonal direction as
$\ell_b \sim \ell/3.5$,
while line-tying will yield a lenght-scale of $\sim L$ for the variation
of $b$ along $z$. We can therefore estimate:
\begin{equation}
\frac{b}{\ell_b} \gtrsim \frac{B_0}{L} \quad \longrightarrow \quad
\frac{b}{B_0} \gtrsim \frac{\ell}{3.5\, L} \label{eq:th}
\end{equation}
With the values used in our simulations ($\ell = 1, L = 10$) this 
rough estimate yields $b/B_0 \gtrsim 3\%$, in agreement
with the simulations presented here, and this is also
approximately the level to which fluctuations settle
in the forced case \citep{rved08}.

\subsection{Ideal simulations}

Current sheet formation if further analyzed with two
ideal simulations of Eqs.~(\ref{eq:eq1})-(\ref{eq:eq2}), with
$\eta_n=\nu_n=0$, $4096^2 \times 2048$ grids, $b_0/B_0 =10\%, B_0=200$,
and two different initial conditions, one with just the mode $m=0$,
and the other one with all modes between 0 and 4 excited, 
with lower modes dominating ($\mathscr{E}_{_m}/\mathscr{E}_{_0} = (m+1)^{-2.6}$).

The \emph{analyticity-strip method} \citep{ssf83, fmb03, bbk13}
extends to the complex space, off the real axis, 
the solutions of ideal MHD equations.
Indicating with $\delta$ the distance from the real 
domain of the nearest complex space singularity, 
this determines in Fourier space an exponential fall-off
at large $k$ for the power-law behavior of the total
energy spectrum (of the real solutions):
\begin{equation}
E(k,t) = C(t)\, k^{-n(t)}\, e^{-2\delta(t) k}.  \label{eq:eqd}
\end{equation}
The width of the strongest current sheet is therefore linked to $\delta$,
and if and how $\delta$ approaches the smallest admissible scale
(fixed at $2$ meshes: $2/k_\textrm{max}, k_\textrm{max}=1364$), determines
whether or not true current sheets form and if the solution develops 
singularities \citep{ssf83, fmb03, kbp11, bm12}.

\begin{figure}
\includegraphics[scale=.4]{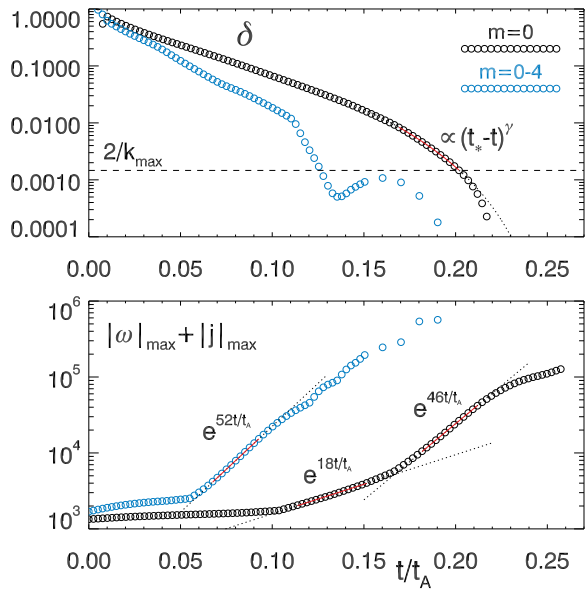}
\caption{Linear-logarithmic plot vs.\ time of the analyticity strip width $\delta$
and of the sum of the current density and vorticity (moduli) maxima,
for the two ideal simulations with only mode $m=0$ and with modes
$m=0$--4 present. The resolution scale is $2/k_\textrm{max}$. 
\label{fig:fig4}}
\end{figure}

For the case with $m=0$, initially $\delta$ decreases exponentially
until time $t\sim 0.16\tau_A$ (Fig.~\ref{fig:fig4}),  after which
it obeys $\delta(t) = C (t_{\ast}-t)^\gamma$ with $t_{\ast} = 0.26\tau_A$
and $\gamma = 3.7$, crossing the resolution scale at $t\sim 0.2\tau_A$,
with a singular-like behavior at $t=t_{\ast}$.
The width $\delta(t)$ is determined fitting the spectrum with
Eq.~(\ref{eq:eqd}), while $t_{\ast}$ and $\gamma$ fit the inverse
logarithmic derivative $\delta/\delta'$ \citep{bbk13}, with a good
linear behavior in this interval.  

The spectral index $n(t) \sim 3.3$, with a maximum value of $\sim 3.7$,
thus $\gamma$ satisfies the condition $\gamma \ge 2/(6-n) \sim 0.87$
that rules out numerical artifacts \citep{bm12, bbk13}.

\begin{figure*}
\includegraphics[scale=.6]{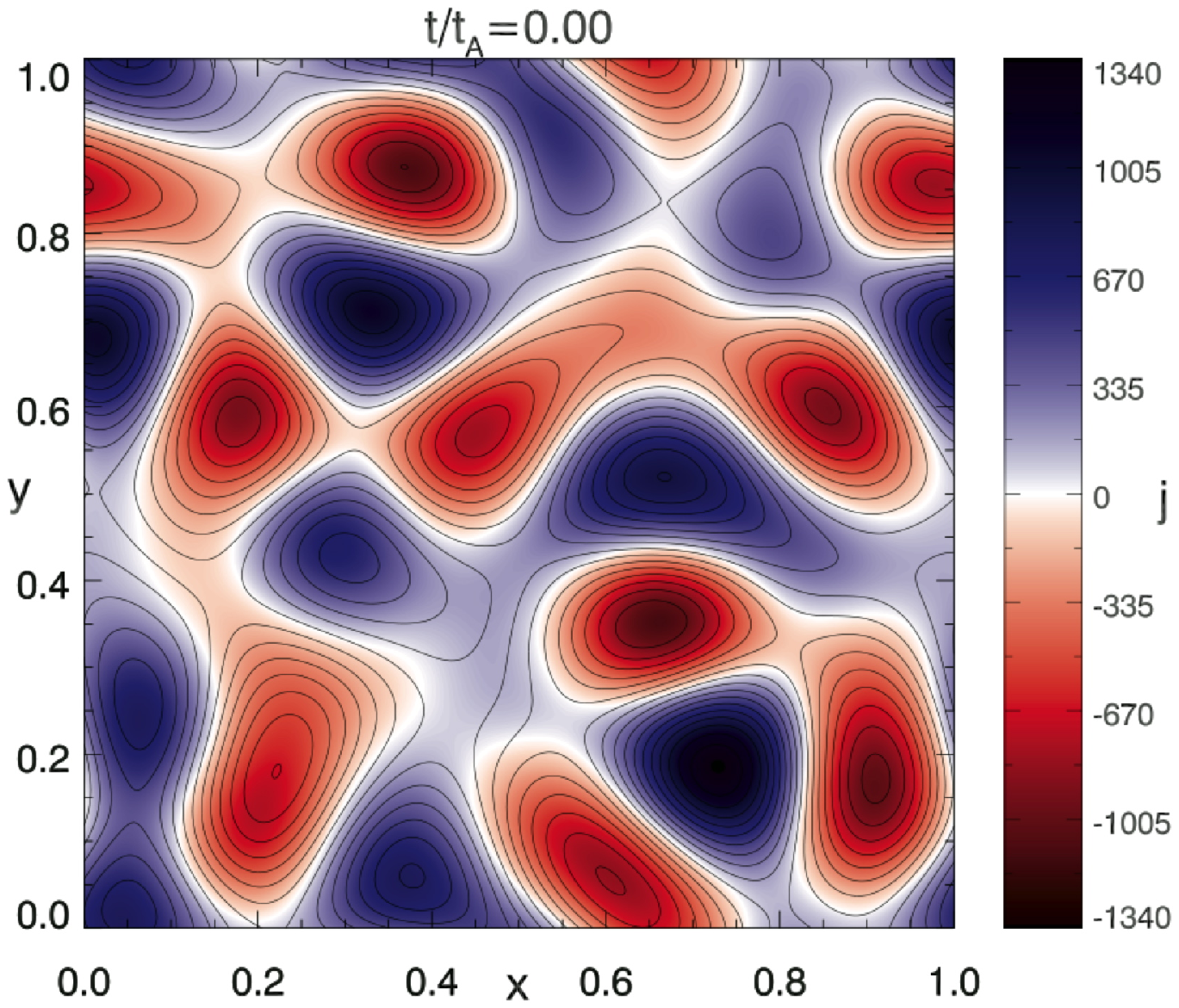}
\includegraphics[scale=.6]{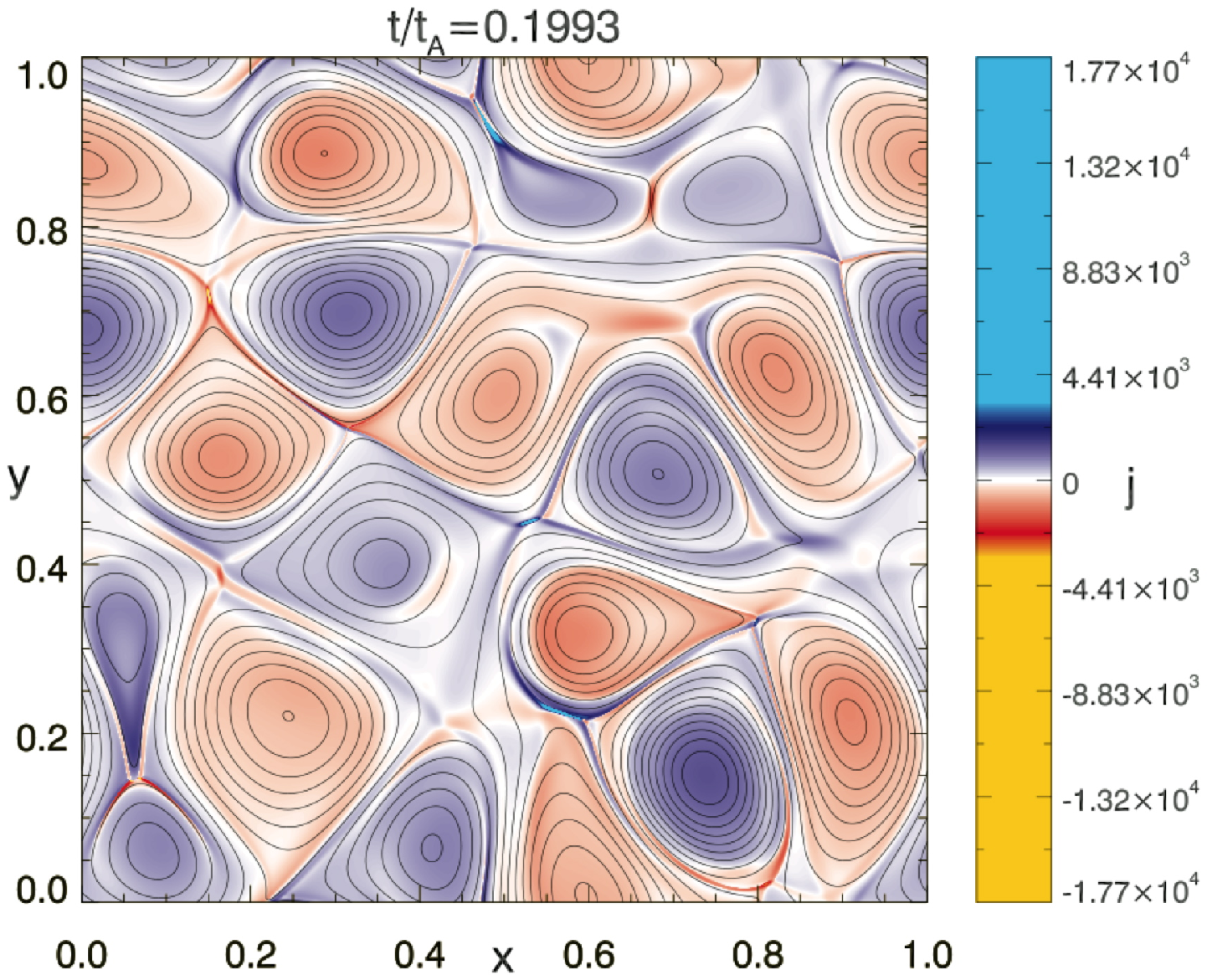}\\
\includegraphics[scale=.6]{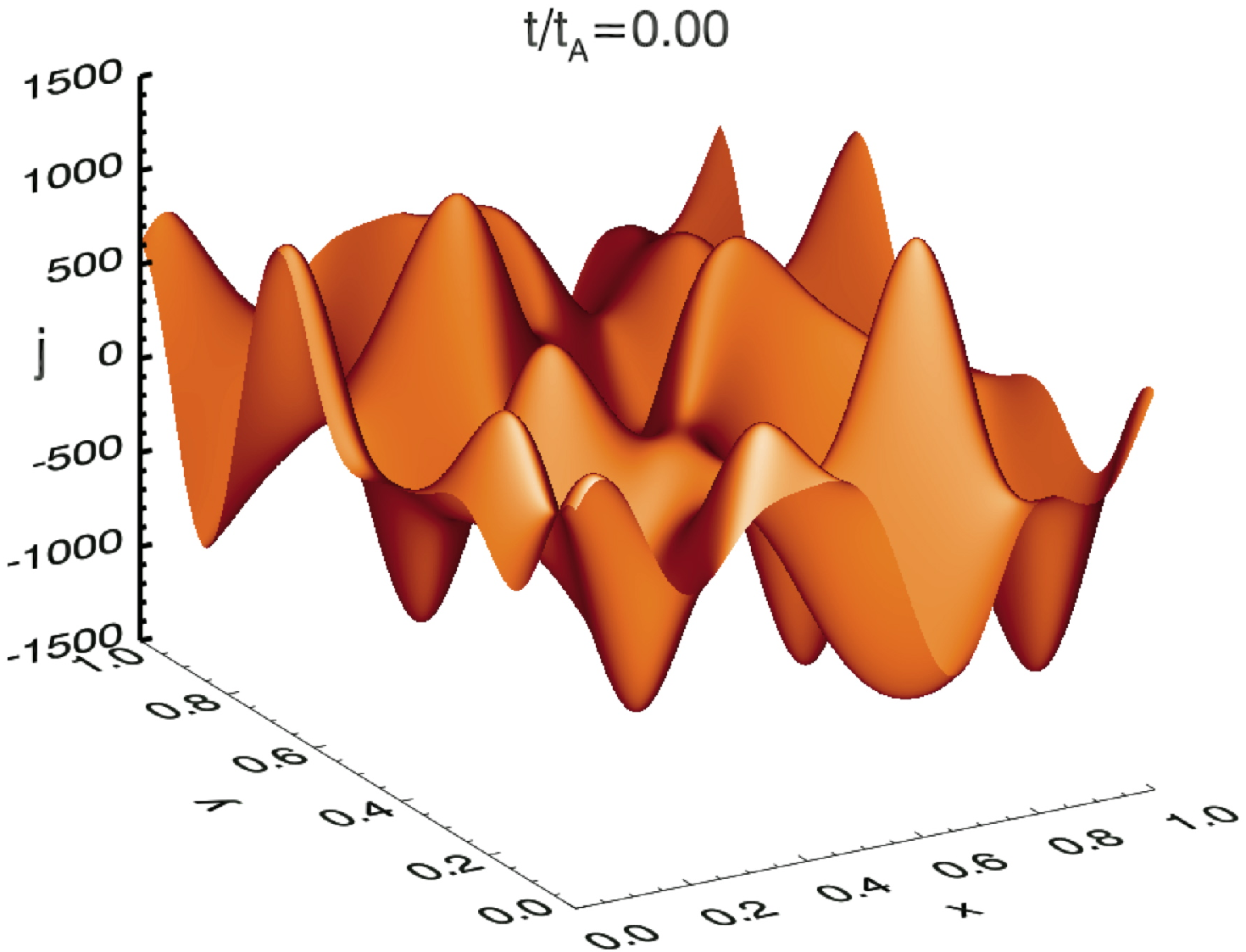}
\includegraphics[scale=.6]{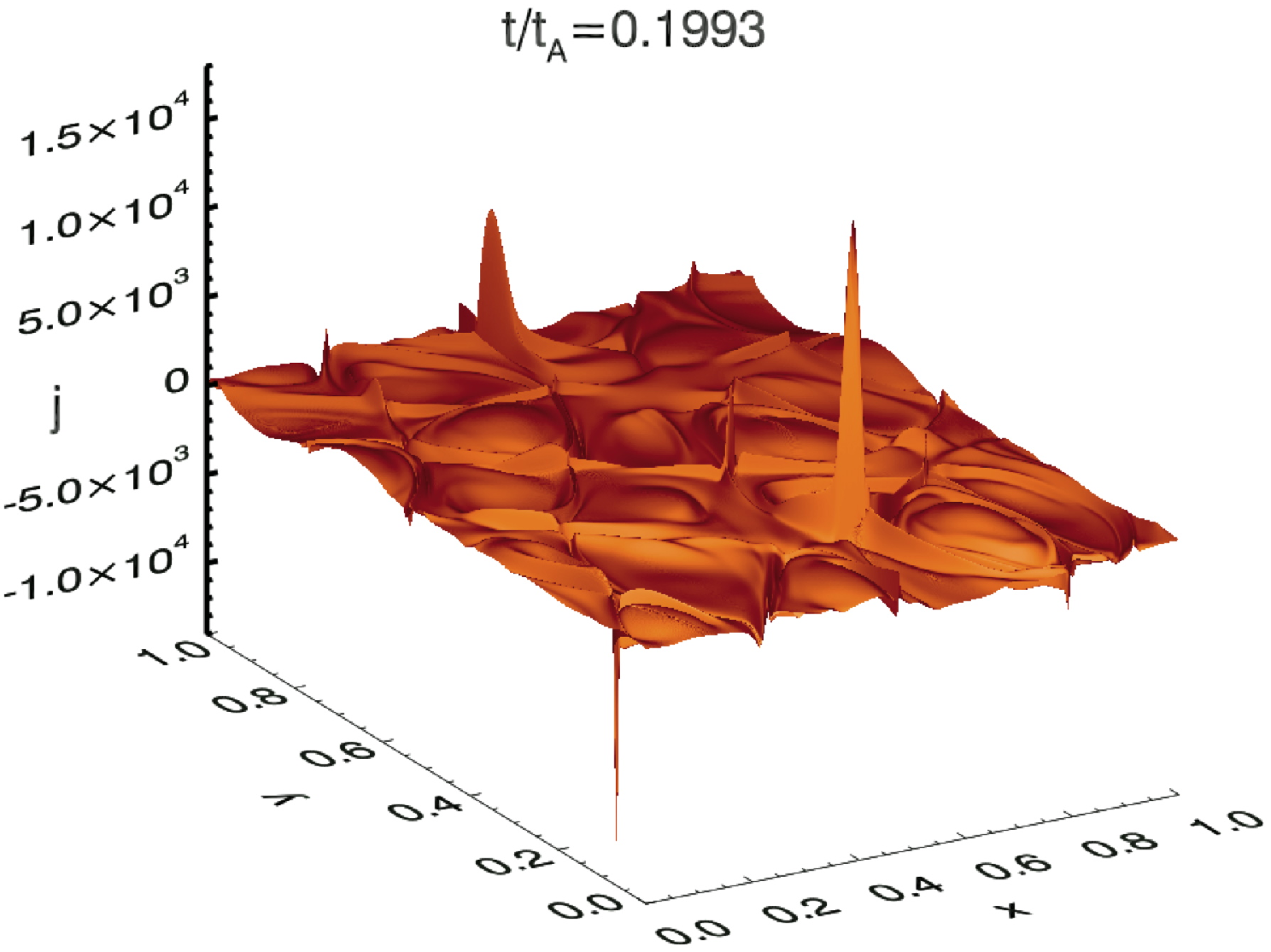}
\caption{Current density $j$ in the mid-plane $z=5$
for the simulation with $m=0$, at $t=0$ (\emph{left column})
and at $t\sim.2\tau_A$ (\emph{right}, 
the color scale reveals thin sheets with strong current enhancements) 
just before $\delta$ crosses
the resolution scale (Fig.~\ref{fig:fig4}). Continuous lines
are field lines of the orthogonal magnetic field.
\label{fig:fig5}}
\end{figure*}

Correspondingly the initially large-scale current density  develops  thin sheets 
with \emph{strong current enhancements} (Fig.~\ref{fig:fig5}) that
approach the resolution scale at  $t\sim .2\tau_A$. Afterward $\delta$
becomes smaller than the mesh size,  the field starts to develop
gaussian statistics \citep{wan09, kbp11}, small-scale noise appears
and the simulation becomes thus under-resolved.

The sum of the suprema of the absolute values of vorticity and 
current (Fig.~\ref{fig:fig4}) exhibits a double exponential behavior
with fast growth rates ($\tau/\tau_A \sim 1/18$ and $1/46$ respectively),
corresponding to two different leading current sheets.
Thus no BKM \citep{bkm84} power-law divergent behavior 
$C(t_\ast -t)^{-\beta}$, with $\beta \ge 1$, is detected.
Therefore one of the three diagnostics for singularities is failed
and a \emph{singular behavior cannot be established}.

When more axial modes are added to the initial condition, 
and the field has a three-dimensional structure,
$\delta$ decreases faster, crossing the resolution threshold
at $t\sim 0.125\tau_A$, and current and vorticity maxima
have a higher exponential growth rate with $\tau/\tau_A \sim 1/52$.

\section{Discussion}

To get insight into the cause of the X-ray
emission of the Sun and  main sequence 
stars we have investigated the dynamical evolution
of magnetic configurations (\ref{eq:pot4})
appropriate to model their coronal fields.

Provided  the value of magnetic
fluctuations $b$ is beyond the threshold (\ref{eq:th}),
we have shown that  current sheets form on fast
ideal timescales, with their thickness 
reaching the resolution scale in the ideal case.
Below this threshold the field-line tension of the line-tied
magnetic field lines inhibits the dynamics and the formation
of current sheets, thus the solutions remain regular. 
As mentioned in the introduction a 
current sheets formation threshold is a critical feature to sustain
an X-ray corona.

The quasi-static dynamics of coronal fields is often modeled as a sequence of instabilities
followed by relaxation and current sheets formation \citep{ng98}, in 
which equilibria play an important role \citep{aly05}.

However the stability and dynamic accessibility of such equilibria 
require further investigations. 
Indeed the majority of these equilibria (Section \ref{sec:eq}) 
do not have the highly symmetric fields required 
for linear instability as, e.g., for kink or other cases \citep{ls93}.
Furthermore the magnetic field induced by  \emph{disordered}  photospheric motions 
is not symmetric, and in general will not be in equilibrium.

In our case the initial magnetic field is not an equilibrium.
In the decaying cases no intermediate equilibria are accessed before current sheets form, 
and no instabilities develop. An approximate (non-symmetric) equilibrium is 
accessed only in the asymptotic regime of the dissipative simulations.
A more complete analysis of the properties of these equilibria and their interplay
with photospheric motions is left to upcoming work.

More in general, the current sheet formation threshold (\ref{eq:th}) might depend
on the specific magnetic topology of the system.

\acknowledgments
A.F.R. thanks Marco Velli for useful discussions.
This research  supported in part by the NASA Heliophysics 
Theory program NNX11AJ44G, 
the NSF Solar Terrestrial and 
SHINE programs AGS-1063439 and AGS-1156094,  
NASA MMS and Solar probe Plus Projects, 
and a subcontract
with the Jet Propulsion Laboratory.
Simulations performed through the NASA Advanced 
Supercomputing SMD award 12-3188.

\end{document}